\documentclass[12pt]{article}

\usepackage{latexsym,amsmath,amssymb,pstricks,wasysym,stmaryrd,enumerate,
epsfig}

\newcommand{\be}{\begin{equation}}
\newcommand{\ee}{\end{equation}}
\newcommand{\beu}{\begin{equation*}}
\newcommand{\eeu}{\end{equation*}}
\newcommand{\bea}{\begin{eqnarray}}
\newcommand{\eea}{\end{eqnarray}}
\newcommand{\beaa}{\begin{eqnarray*}}
\newcommand{\eeaa}{\end{eqnarray*}}
\newcommand{\bmx}{\begin{pmatrix}}
\newcommand{\emx}{\end{pmatrix}}

\newcommand{\del}{\partial}
\newcommand{\g}{{\frak g}}
\newcommand{\h}{{\frak h}}
\newcommand{\m}{{\frak m}}

\newcommand{\D}{{\cal D}}

\newcommand{\Pf}{{\rm Pf}}

\advance\textheight by \topskip
\begin{document}
\baselineskip 16pt
\parindent 8pt
\parskip 5pt

\begin{flushright}
DAMTP-2004-84
\break
hep-th/0408244v2\\[3mm]
\end{flushright}

\begin{center}
{\Large {\bf Quantum, higher-spin, local charges \\
in symmetric space sigma models}}\\
\vspace{0.8cm} 
{\large J.M. Evans${}^a$,
D. Kagan${}^a$,
N.J. MacKay${}^b$,
C.A.S. Young${}^a$,
}
\\
\vskip 5pt
{\em ${}^a$DAMTP, Centre for Mathematical Sciences, University of
Cambridge,\\ Wilberforce Road, Cambridge CB3 0WA, UK}
\\
\vskip 5pt
{\em ${}^b$Department of Mathematics, University of York,\\
Heslington Lane, York YO10 5DD, UK}
\\
\vskip 5pt
{\small E-mail: {\tt J.M.Evans@damtp.cam.ac.uk}, 
{\tt D.Kagan@damtp.cam.ac.uk}, \\ {\tt nm15@york.ac.uk},
{\tt C.A.S.Young@damtp.cam.ac.uk}}
\\
\end{center}

\vskip 0.15in
 \centerline{\small\bf ABSTRACT}
\centerline{
\parbox[t]{5in}{\small
Potential anomalies are analysed for the local spin-3 and spin-4
classically conserved currents in any two-dimensional sigma model
on a compact symmetric space $G/H$, with $G$ and $H$ classical 
groups. Quantum local conserved charges are shown to exist in exactly those 
models which also possess quantum non-local (Yangian) charges.
The possibility of larger sets of quantum local charges is discussed 
and shown to be consistent with known S-matrix results and the 
behaviour of the corresponding Yangian representations.
}}

\vspace{20pt}
\centerline{\bf 1. Introduction}

Sigma models in two spacetime dimensions are known to be quantum integrable 
if their target manifolds are compact Lie groups or certain other 
symmetric spaces \cite{Luscher}-\cite{GW}, namely
\bea
&SO(n{+}1)/SO(n) \ , \qquad SU(n)/SO(n) \ , \qquad SU(2n)/Sp(n),
\nonumber\\
&SO(2n)/SO(n){\times}SO(n) \ , \qquad 
Sp(2n)/Sp(n){\times}Sp(n) \ ,& 
\label{list} 
\eea
together with a finite number of examples involving exceptional groups.
It is a long-standing result \cite{AFG} that a sigma model on a
symmetric space $G/H$ with $H$ \emph{simple}\footnote{Here, and throughout
this paper, we shall use `simple' to mean that the corresponding Lie 
algebra has no non-trivial ideals. 
Hence $U(1)$ is simple in our terminology, in addition to the usual
non-abelian simple groups of the Cartan-Killing 
classification \cite{Helgason}.} possesses 
a conserved \emph{non-local} charge at the quantum level, 
which implies integrability.
The real and symplectic Grassmannians in (\ref{list}) 
have $H$ \emph{non-simple}, however, and are comparatively recent 
additions to this list of integrable models.
S-matrices for these new cases were proposed and tested 
in \cite{Fendley:2000,Fendley:2001,Babichenko} 
and it was subsequently shown \cite{EKY}
that they too possess quantum non-local charges. 

An alternative criterion for integrability, either classical or 
quantum, is the existence of higher-spin, \emph{local} conserved
charges (see e.g.~\cite{Parke,GW,EFlocal,PCM,Evans00,Evans01}). 
These offer an independent check on whether the list in 
(\ref{list}) is now correct and complete, which is particularly important 
in view of the fact that two families were overlooked until recently.
In this paper we give a uniform treatment of sigma models on
(irreducible \cite{Helgason}) symmetric spaces $G/H$ 
where $G$ (simple) and $H$ are compact classical groups:
we investigate the quantum behaviour of all local conserved currents
of spin three or four using an approach due to 
Goldschmidt and Witten \cite{GW} which is based purely on 
symmetry arguments.\footnote{Goldschmidt and Witten \cite{GW}
established quantum integrability for sigma models on spheres and 
classical Lie groups (see also \cite{PCM});
the method was also used in \cite{EKY}, but only for a rather 
special set of currents, to confirm the integrability of the 
$SO(2n)/SO(n){\times}SO(n)$ models (see section 5).}
 
We discuss general aspects of the classical symmetric space sigma models 
in section 2, before outlining how the analysis of the local
currents is carried out, and also stating our results, in section 3.
Sections 4 and 5 provide the case-by-case details of the analysis on 
which the results rest. 
Our findings are in complete agreement with the
studies of non-local charges cited above. In addition to the 
basic question of integrability of the models, however, there are also very 
interesting consequences which follow from the co-existence of local 
and non-local conserved quantities. 
Both constrain the S-matrix in powerful ways 
(beyond factorisation) and the consistency of these constraints is 
highly non-trivial \cite{dorey91,Corrigan,chari95,PCM}.
We discuss, in section 6, how the existence of 
quantum local charges fits with the properties of the Yangian 
quantum groups generated by the non-local charges 
(see e.g.~\cite{Bernard,MacKay92,EF,AABZ}).

\vspace{15pt}
\centerline{\bf 2. General aspects of the models}

One way to formulate the sigma model with target space
$G/H$ is to introduce a field $g(x^\mu) \in G$ which transforms under 
both a global symmetry $g (x^\mu) \mapsto U g(x^\mu)$, for $U \in G$,
and also under local gauge transformations $g (x^\mu) \mapsto g
(x^\mu) h(x^\mu)$ for $h(x^\mu) \in H$, thereby ensuring that 
physical degrees of freedom are properly confined to the coset
space.\footnote{All groups are classical and elements of them are
unitary matrices in the defining representations; similarly,
elements of the corresponding Lie algebras are traceless antihermitian 
matrices.}
An alternative approach, which we shall adopt, is to construct the
model without any gauge redundancy.
To do this we use the fact that 
each symmetric space can be parametrised by a particular 
set of unitary matrices, so there is a field $\Phi_{ab}(x^\mu)$ 
obeying $\Phi \Phi^\dagger =1$ as well as certain additional 
constraints. We will describe the precise nature of these 
constraints for each family of classical symmetric spaces in 
sections 4 and 5; in this section we will concentrate 
on features common to all the models.

We must specify how $\Phi$ transforms under $G$, and we will distinguish 
two possibilities:
\be 
{\rm (a)} \quad \Phi \mapsto U \Phi U^\dagger \qquad {\rm or}\qquad
{\rm (b)} \quad \Phi \mapsto U \Phi U^T \label{sym} 
\ee
where $U \in G$. The constraints on $\Phi$ ensure that
it can be related to a \emph{canonical form}, by which we mean some fixed, 
unitary matrix $N$, as follows:
\be 
{\rm (a)} \quad \Phi = g N g^\dagger \qquad {\rm or} \qquad
{\rm (b)} \quad \Phi = g N g^T \ . \label{form} 
\ee
Any given $\Phi (x^\mu)$ thus corresponds to (many) $g(x^\mu)$,
and this is exactly the field introduced in the coset 
description. The correspondence between the two approaches 
is completed by identifying $H$ as the subgroup which 
preserves $N$:
\be 
{\rm (a)} \quad h N h^\dagger = N \qquad {\rm or} \qquad
{\rm (b)} \quad h N h^T = N \ . \label{hinv} 
\ee

Note that cases (a) and (b) coincide if $G$ is an orthogonal group.
Case (b) applies to the families $SU(n)/SO(n)$ and $SU(2n)/Sp(n)$,
while case (a) applies to all other classical symmetric spaces. 
For example, if $\Phi$ is a traceless, hermitian, $2n{\times}2n$ matrix
transforming under $G = SU(2n)$ as in (\ref{sym}{a}), then 
(\ref{form}{a}) certainly holds with 
$N = {\rm diag} (1, \ldots , 1, -1, \ldots, -1)$ and 
(\ref{hinv}{a}) implies that $H=S(U(n){\times}U(n))$.
In this way we recover one of the complex Grassmannians.
On the other hand, the coset $SU(n)/SO(n)$ is obtained by 
taking $\Phi$ to be a complex, symmetric, $n{\times}n$ matrix with 
$\det \Phi = 1$. It transforms under $SU(n)$ according to (\ref{sym}{b}),
so (\ref{form}{b}) holds with $N$ the identity matrix 
and the subgroup $H$ is indeed $SO(n)$.
 
Underlying this construction, in general, is the Cartan immersion  
$G/H \rightarrow G$ \cite{EF,CI}. But the validity of the
approach can be checked case by case when $G$ and $H$ are classical, 
using basic results from linear algebra, as in the examples above
(and as in \cite{Fendley:2000,Fendley:2001,MS}).  
When we consider the various families of symmetric spaces in more
detail in sections 4 and 5, we will simply state the properties 
required of $\Phi$ and specify the canonical form $N$.

The lagrangian for the $G/H$ sigma model,
with field $\Phi (x^\mu)$ is 
${\rm Tr} ( \del_\mu \Phi \del^\mu \Phi^\dagger)$
supplemented by the relevant constraints,
among them $\Phi \Phi^\dagger =1$ (these can be enforced 
by Lagrange multipliers). 
The theory is invariant under $G$, with Noether current 
\be
j_\mu \, = \, {\textstyle \frac{1}{2}} \, \Phi \, \del_\mu
\Phi^\dagger \ \in \g \ ,
\label{ncurr}\ee
where $\g$ is the Lie algebra of $G$. This current transforms
\be
j_\mu \mapsto U j_\mu U^\dagger
\ee
under (\ref{sym}) in cases (a) \emph{and} (b).
The equations of motion for the model are\footnote{Orthonormal and
light-cone components of spacetime vectors are related 
by $u_\pm = u_0 \pm u_1$.}
\be
\del_- j_+  \, = \,
- \del_+ j_-  \, = \,
[ j_+  , j_- ]
\label{lax}\ee
or in terms of the fields,
\be 
\del_+ \del_- \Phi = 2 \{ j_+ , j_- \} \Phi \ .
\label{eqnmot}\ee
Note that this is independent of the details of the constraints.

The form of equation (\ref{lax}) is responsible for the classical
integrability of the symmetric space sigma models. It allows
us to construct local conserved currents of spin $m$:
\be 
\del_- {\rm Tr} (j_+^m) = 0
\label{curr}\ee
(and similarly with $\pm$ interchanged).
It is, of course, essential to know when such currents are non-zero,
and this leads us to the question of \emph{invariants}
on symmetric spaces.

Let $\h$ be the Lie algebra of $H$ and 
$\m$ its orthogonal complement in $\g$; then
\be \g=\h \oplus \m  \qquad {\rm with} \qquad 
[\h,\h]\subset\h \ ,\quad [\h,\m]\subset\m \ , \quad[\m,\m]\subset\h \
.
\label{symm}\ee
It follows from the definition of $H$ 
in (\ref{hinv}a) or (\ref{hinv}b) 
that $\h$ and $\m$ are the $\pm 1$ eigenspaces
of the following map on $\g$ 
\be
{\rm (a)}  \quad \sigma: X \mapsto N X N^\dagger \qquad 
{\rm or} \qquad {\rm (b)} \quad
\sigma : X \mapsto N X^* \! N^\dagger \ .
\ee
In other words, $\sigma$ is the involutive automorphism of $\g$ which 
defines the symmetric space.

Now although $j_\mu$ belongs to $\g$, it is conjugate to something
in the subspace $\m$, specifically
\be j_\mu = - g k_\mu g^{-1}  \qquad {\rm where} \qquad k_\mu = 
{\textstyle{\frac{1}{2}}} 
( \, g^{-1} \del_\mu g  -   \sigma(g^{-1} \del_\mu g) \, )  \in \m
\label{conj}\ee
which follows on substituting (\ref{form}) in (\ref{ncurr}). 
In considering ${\rm Tr} (j_+^m) = (-1)^m {\rm Tr} (k_+^m)$ 
we are therefore concerned not merely with symmetric $G$-invariant 
tensors on $\g$, but actually  
with symmetric $H$-invariant tensors on $\m$. We shall refer 
simply to \emph{invariants on} $G/H$ from now on.
An analysis of these symmetric $G/H$ invariants, including the question of 
which of them are \emph{primitive}, meaning that they are not combinations of 
invariants of lower order, has been given in \cite{Evans00} 
(see also \cite{Evans01}) in the context of classical conserved
charges; we shall have frequent recourse to these results. 

Discrete symmetries of the $G/H$ sigma models also play a very
important role. All models are invariant under 
\be 
\Phi \mapsto \Phi^* \, , \qquad j_\mu \mapsto j_\mu^* = -j_\mu^T \ .
\label{cc}\ee
Even if this symmetry is trivial, in that $\Phi$ is real,
the resulting antisymmetry of the current has important consequences, 
e.g.~for the vanishing of certain trace invariants. Other 
discrete symmetries will be discussed as they arise.

\vspace{15pt}
\centerline{\bf 3. Overview of anomaly counting and statement of results}

Suppose that in some model there are $n_C$ independent, local quantities 
$C_k$ which obey classical conservation equations $\del_- C_k = 0$ and 
which have identical behaviour under all symmetries, continuous and discrete. 
The Goldschmidt-Witten (GW) approach \cite{GW} to investigating the effect of
quantization is to count the number of independent, local terms $A_i$
with the correct symmetries to appear as quantum modifications of 
these equations, making free use of the classical equations of motion,
and to count similarly the number of independent terms $B_j$ which
have the correct symmetries \emph{and} which are derivatives of local 
expressions. Let the number of these terms be $n_A$ and $n_B$ respectively.
If $n_C > n_A - n_B$ then there is at least one combination of
conservation equations which survives quantization,
because its quantum modification is a derivative. 
When this occurs for a given class of conservation equations
(characterised by their symmetries, including spin) 
we will say `GW \emph {works}', and if not, `GW \emph{fails}'.
Note, however, that `GW fails' does not imply that there is no quantum
conservation equation, only that we can draw no definite
conclusion from the counting.

We will investigate the 
quantum modifications to (\ref{curr}) 
with $m=2, 3, 4$. The anomalies ($A$'s) and derivatives ($B$'s) must be
$G$-invariant and so can be expressed in terms of\footnote{When $G$ is
$SO(n)$ or $SU(n)$ we can also construct invariants using
$\varepsilon$ tensors. These are never relevant for the anomalies and
derivatives associated with currents (\ref{curr}), however, 
because the $\varepsilon$ invariants either reduce to traces or else
can be distinguished from trace-type invariants by symmetries. 
Some details are given in sections 4 and 5.} 
\be 
{\rm (a)} \quad {\rm Tr} ( \, \D_1 \Phi^{\phantom{\dagger}} 
\D_2 \Phi \ldots \D_r \Phi^{\phantom{\dagger}}  )
\qquad {\rm or} \qquad 
{\rm (b)} \quad {\rm Tr} ( \, \D_1 \Phi^{\phantom{\dagger}} \D_2
\Phi^\dagger \ldots \, \D_{2s-1} \Phi^{\phantom{\dagger}} 
\D_{2s} \Phi^\dagger ) 
\label{ginv} 
\ee
depending on the transformation in (\ref{sym}), where each 
$\D_i$ is a product of powers of $\del_+$ and $\del_-$, or the
identity operator. But it follows 
from $\Phi \Phi^\dagger =1$ and consequences of it such as 
$(\del_\mu \Phi) \Phi^\dagger 
= - \Phi (\del_\mu \Phi)^\dagger$ that all derivatives of $\Phi$ can 
be re-expressed in terms of the Noether current $j_\mu$, its
derivatives, and $\Phi$ fields \emph{without} derivatives. 
Moreover, any quantity of the form (\ref{ginv}b), or (\ref{ginv}a) 
with $r$ even, can be written \emph{entirely} in terms of $j_\mu$ and 
its derivatives, 
while any quantity (\ref{ginv}a) with $r$ odd can be written in
terms of $j_\mu$, its derivatives, and a \emph{single} $\Phi$.
Finally, we can use (\ref{lax}) to eliminate 
any occurrences of $\del_- j_+$ or $\del_+ j_-$.

The simplest application of GW counting is to energy-momentum
conservation. The classical equation $\del_- {\rm Tr} (j_+^2) = 0$ will
be modified in the quantum theory (scale-invariance is broken)
but we can check, in each model (see sections 4 and 5), that the 
only relevant anomaly terms with the correct symmetries 
are those involving just $j_\mu$ and its derivatives.
The quantum modification must be even under (\ref{cc}) and so must be 
proportional to 
${\rm Tr} (j_- \del_+ j_+) = \del_+ {\rm Tr} (j_- j_+)$,
on using the equations of motion. Thus energy-momentum is conserved 
quantum-mechanically, as expected.

Consider next the classical spin-3 current,
$C = {\rm Tr} (j_+^3)$, which is odd under (\ref{cc}). 
For this to be non-vanishing, the symmetric space must have 
a 3rd-order invariant, which occurs only for 
$SU(n)/SO(n)$ and $SU(2n)/Sp(n)$ with $n\geq 3$ \cite{Evans00}.
But in these families the field $\Phi$ transforms as in (\ref{sym}b),
so $G$-invariants are of type (\ref{ginv}b) and, as argued above, 
anomalies and derivatives can once again be written entirely in terms 
of $j_\mu$ and its derivatives.
The only terms with the correct symmetries, changing sign under
(\ref{cc}), are 
\be
A = {\rm Tr} (j_- \{ j_{+}, \del_+ j_{+} \}) \ , \qquad B = 
\del_+ {\rm Tr} (j_- j^2_{+}) \ .
\ee 
Hence GW works (in fact $A=B$) and the spin-3 current 
survives quantization.

Now consider the conserved quantities of spin-4:
\be
C_1 = ( \, {\rm Tr} (j_+^2) \, )^2 \ , 
\quad
C_2 = {\rm Tr} (j_+^4) \ ,  
\label{spin4}\ee 
which are both invariant under (\ref{cc}).
These are distinct if there is a primitive symmetric 4th-order 
invariant on $G/H$; if there is not, $C_2$ is proportional 
to $C_1$. Thus $n_C = \mbox{1 or 2}$, \emph{the number of 
independent symmetric 4th-order invariants on} $G/H$. 

Before listing possible anomalies and derivatives for the
spin-4 case it is helpful to introduce some new notation by defining:
\be
j_{++} = \del_+ j_+ \ , \quad 
j_{+++} = \del_+ j_{++} \, + \, [ j_+ , j_{++} ] \ , \quad
j_{++++} = \del_+ j_{+++} \, + \, [ j_+ , j_{+++} ] \ .
\label{pluses}\ee
The advantage of working with these modified derivatives, all of 
which clearly belong to $\g$, is that they each obey 
$j = - g k g^{-1}$ for some 
$k \in \m$, generalising (\ref{conj}).
This, and properties which follow from it, will
prove very convenient. Note also 
$j \mapsto j^* = -j^T$ under (\ref{cc}).

The counting of anomalies and derivatives is much more 
intricate for the spin-4 case. Let us first write down
all possible anomaly terms which are constructed from $j_\mu$ alone
and which are invariant under $(\ref{cc})$. 
There are five independent terms in general:
\bea
A_1 & = & {\rm Tr} (j_- j_{++++}) \nonumber \\
A_2 & = & {\rm Tr} (j_- j_{+}) {\rm Tr} (j_+ j_{++}) \nonumber \\
A_3 & = & {\rm Tr} (j_- j_{++}) {\rm Tr} (j^2_+ ) \nonumber \\
A_4 & = & {\rm Tr} (j_+ j_- j_+ j_{++}) \nonumber \\
A_5 & = & {\rm Tr} (j^2_+ \{ j_- , j_{++} \}) \label{basicA}
\eea
Other possibilities which involve $\Phi$ fields and which we 
encounter in some models are:
\bea
A_6 & = & {\rm Tr} ( \Phi j_- j_{+}) {\rm Tr} ( \Phi j_+ j_{++})
\nonumber \\
A_7 & = & {\rm Tr} ( \, \Phi j^2_+ \{ \, j_- , j_{++} \} \, ) \label{extraA} 
\eea
The derivative terms which involve $j_\mu$ alone 
and which have the correct symmetries are:
\bea
B_1 & = & \del_+ {\rm Tr} (j_- j_{+++}) \nonumber\\
B_2 & = & \del_- {\rm Tr} ( j_{++}^2) \nonumber\\
B_3 & = & \del_+ ( \, {\rm Tr} (j_- j_{+}) {\rm Tr} (j^2_+ ) \, ) \nonumber\\
B_4 & = & \del_+ {\rm Tr} (j_- j_+^3) \label{basicB}
\eea
It will turn out that these are the only derivative terms that
are relevant. (Note that $\del_- {\rm Tr} (j_+ j_{+++})$ is proportional
to $B_2$, using the equations of motion.)

It is clear that $B_1$, $B_2$, $B_3$ are always
independent, but $B_4$ is an additional independent quantity iff
there is a primitive 4th-order invariant on $G/H$, otherwise it is
proportional to $B_3$. Thus  
\be
n_C = 1 \ \Rightarrow \ n_B = 3 \ , \qquad
n_C = 2 \ \Rightarrow \ n_B = 4 \ .
\ee
The question of how the various anomaly terms are related when $n_C =
1$ is more subtle for two reasons.
First, the lack of a primitive 4th-order invariant implies only that
\emph{totally symmetrized} traces reduce to lower-order invariants.
For instance, the appropriate combination of $A_4$ and $A_5$ in (\ref{basicA}) 
must reduce in these circumstances, but they 
need not be individually reducible.
Second, $G$-invariant quantities simply work differently if they involve
$\Phi$, such as $A_6$ and $A_7$ in (\ref{extraA}), because, 
unlike the currents, $\Phi$ does not belong to $\g$. 
These cases require special consideration (one approach is explained
in the appendix).

We have kept the presentation as general as possible for as long as
possible; to go further we must state explicitly 
the constraints obeyed by $\Phi$, the discrete symmetries 
of the sigma model, and so on.
We provide these details in the following sections, 4 and 5, 
and we specify exactly which anomaly and derivative terms are allowed
and independent for each classical symmetric space. 
In the interests of clarity, however, we state, in advance,
the conclusions which follow from GW counting:

\noindent
{\bf (i)} \emph{All sigma models in} (1) \emph{possess at least one 
higher-spin quantum conserved charge.} 

\noindent
{\bf (ii)} \emph{GW fails for sigma models 
on those classical symmetric spaces $G/H$ 
not in} (1). 

\noindent
These broad conclusions provide independent confirmation of
the quantum integrability of the models in (1) and also lend support to the 
suggestion that this list is complete. At a finer level of detail,
however, we find:

\noindent
{\bf(iii)} \emph{The sigma models on $SU(n)/SO(n)$ and $SU(2n)/Sp(n)$
possess a spin-3 quantum conserved current for $n \geq 3$
(these are the classical symmetric spaces with a 
3rd-order invariant).}

\noindent
{\bf (iv)} \emph{All sigma models in} (1) 
\emph{possess a spin-4 quantum conserved current, with the 
possible exceptions of $SU(3)/SO(3)$ and 
$SU(6)/Sp(3)$, for which GW fails.}

\noindent 
We will also summarise, and clarify in one small respect,
the status of local charges in principal chiral models (PCMs) with 
target spaces simple classical Lie groups $G$ \cite{GW,PCM}:

\noindent
{\bf (v)} \emph{There is a quantum spin-3 conserved current when $G$ is
$SU(n)$ for $n \geq 3$ (these are the classical groups with 
a 3rd-order invariant); and there is a quantum spin-4 conserved  
current for all $G$, with the possible exception of $SU(3)$, for which 
GW fails.}

\noindent
The analysis of spin-4 currents in PCMs was given in \cite{GW,PCM}
for classical groups $G$ which possess a primitive 4th-order invariant. 
Of the few classical groups which lack such an invariant, the case 
$SU(2) = S^3$ was also dealt with in \cite{GW}, as one of the family 
of sigma models on spheres, and $SO(3)$ is similar. 
This leaves just the case of $SU(3)$ to consider, which is easily done 
(see section 5).

It is striking that there are just three examples---in (iv) and 
(v)---where the sigma model is integrable yet GW fails to show the 
quantum conservation of a spin-4 current (even though it shows the
existence of a quantum spin-3 current). 
We shall return to this point in section 6.

\vspace{15pt}
\centerline{\bf 4. Details for the Grassmannian models}

The real, complex and quaternionic (or symplectic) Grassmannians:
\be SO(p{+}q)/SO(p){\times}SO(q) \ , \quad
SU(p{+}q)/S(\, U(p){\times}U(q) \, ) \ , \quad
Sp(p{+}q)/Sp(p){\times}Sp(q) 
\nonumber \ee
are each parametrised by a hermitian matrix $\Phi$ and 
(\ref{sym}a,\ref{form}a,\ref{hinv}a) hold.
In the real and complex cases $\Phi$ is $(p{+}q){\times}(p{+}q)$ 
with\footnote{
There is a subtlety in the real case: 
the matrix $\Phi$ actually parametrises
$O(p{+}q)/O(p){\times}O(q) = SO(p{+}q)/S(\, O(p){\times}O(q) \,)$ 
rather than $SO(p{+}q)/SO(p){\times}SO(q)$  
and so it does not quite provide a `faithful
representation' of the latter symmetric space. The relationship 
between these spaces is analogous to that between $SU(2)$ and $SO(3)$
(which is actually the case $p=3, q=1$).
}
\be 
\Phi^\dagger = \Phi \ , \qquad \Phi^2 =1 \ , \qquad 
{\rm Tr} \, \Phi = p-q \ , \qquad N = 
\bmx 1_{p \times p} & 0 \\  0      & -1_{q \times q }  \emx \ .
\ee
In the quaternionic case $\Phi$ and $N$ are doubled in
size, with ${\rm Tr} \Phi = 2(p{-}q)$. For the real and 
quaternionic families $\Phi$ also satisfies reality conditions
\be \Phi^* = \Phi  \qquad {\rm or} \qquad \Phi^* = J \Phi J^{-1} \ , 
\label{real}\ee
respectively, where $J$ is a $2(p{+}q){\times}2(p{+}q)$ symplectic
structure (a real antisymmetric matrix with $J^2=-1$).
All three families of Grassmannians have a special feature 
which arises iff $p=q$, namely, a discrete symmetry
$ \tau: \Phi \mapsto - \Phi$, $j \mapsto j$, where $j$ is 
any of the currents $j_\pm$ or $j_{+ \ldots +}$.
(Note that if $p \neq q$ this map does not respect the condition
on ${\rm Tr} \Phi$.)
The conserved quantities (\ref{curr}) are clearly invariant under $\tau$.
This symmetry was crucial in showing the existence of 
quantum conserved non-local charges for the special families of
Grassmannians in (\ref{list}) \cite {EKY} and it will play an equally 
important role here.

Using $\Phi = \Phi^\dagger$ and the definitions (\ref{ncurr}) and
(\ref{pluses}) we find, for any current $j$, 
\be
\Phi j = - j \Phi  \qquad {\rm or } \qquad \Phi j \Phi^{-1} = -j \ ,
\label{anti}\ee 
and this has useful consequences: (i) The trace of any odd number of 
currents vanishes, whether or not it contains a factor of $\Phi$.
(ii) ${\rm Tr} (\Phi j^2) = 0$ for any $j$, from cyclicity of 
the trace. (iii) ${\rm Tr} (\Phi j j')$ is odd under (\ref{cc}); in the real 
and symplectic cases (\ref{real}) then implies that it actually vanishes;
in the complex case it need not vanish if $j \neq j'$, 
but we must multiply two such expressions to obtain
something even under (\ref{cc}), an example being $A_6$ in 
(\ref{extraA}). (iv) Similar restrictions can
be derived for traces of $\Phi$ with four currents; thus 
$A_7$ in (\ref{extraA}) is non-zero 
and even under (\ref{cc}), whereas other candidates for anomalies
and derivatives, such as ${\rm Tr} (\Phi j_+ j_- j_+ j_{++})$ and 
$\del_\mp {\rm Tr} (\Phi j_\pm j^3_+)$, either vanish or are odd
under (\ref{cc}).

These remarks justify the assumption made in section 3 
(of using only $j$'s, no $\Phi$'s) when establishing 
quantum conservation of energy-momentum. 
They also imply that for spin-4 currents, the 
possible anomalies are limited to $A_i$ in (\ref{basicA}) and (\ref{extraA})
while the only derivatives with the correct symmetries are 
$B_j$ in (\ref{basicB}).
We now complete the treatment of the spin-4 case 
for each family. (Some additional comments,  mainly of use in confirming 
the counting in the $q=1$ cases, are relegated to an appendix.)

\noindent
\emph{Spin-4 for real Grassmannians}
\hfill \break
$\bullet$ If $p \geq q \geq 2$ then $n_C = 2$ and $n_B=4$.
The possible anomalies are $A_i$ with $1 \leq i \leq 5$ 
and $A_7$ if $p \neq q$. If $p = q$, however, then $A_7$ 
is ruled out by $\tau$. 
Thus for $p \neq q$ we have $n_A = 6$, $n_B = 4$ and GW fails;
but for $p=q$ we have $n_A = 5$, $n_B = 4$ and GW works.
\hfill \break
$\bullet$ If $p > q=1$, the target manifolds are spheres $S^p$.
There is no primitive 4th-order invariant so $n_C = 1$ and $n_B=3$. 
But $A_4$, $A_5$ and $A_7$ can now each be expressed in terms 
of $A_2$ and $A_3$. Hence $n_A = n_B = 3$ and GW works for all $p$. 

We have not considered invariants involving $\varepsilon$
tensors above, because the real Grassmannians possess 
a symmetry  $\mu : \Phi \mapsto M \Phi M^T$, where $M$ is an
orthogonal matrix with $\det M = -1$. 
Trace-type invariants, and so all the higher-spin currents, 
anomalies and derivatives, are inert under $\mu$. 
But an invariant constructed from a single $\varepsilon$ tensor 
changes by a factor $\det M = -1$ under $\mu$.
Precisely this property was used in \cite{EKY} to show 
that the conservation of the classical Pfaffian\footnote{
The Pfaffian is defined by $\Pf X = c_n \, 
\varepsilon_{a_1 b_1 a_2 b_2 \dots a_n b_n} \, 
X_{a_1 b_1} X_{a_2 b_2} \dots X_{a_n b_n}$, 
for any $2n{\times}2n$ antisymmetric matrix $X$ 
(real or complex), where the constant 
$c_n = (2^n n!)^{-1}$ ensures $(\Pf X)^2 = \det X$. 
}
current $\del_- {\rm Pf}
(j_+) = 0$, which exists iff $p=q$, 
generalises to the quantum theory. The anomaly must be odd under
$\mu$ but even under $\tau$, which leaves 
$ \del_+ ( \, \varepsilon_{a_1 b_1 a_2 b_2 \dots a_p b_p} \, 
j_-^{a_1 b_1} j_+^{a_2 b_2} \dots j_+^{a_p b_p} \, ) $
as the only possibility. Hence the $SO(2p)/SO(p){\times}SO(p)$ 
models contain both a quantum current of spin 4 which is even under
$\mu$ and a quantum current of spin $p$ which is odd under $\mu$.

\noindent
\emph{Spin-4 for complex Grassmannians} \hfill \break
$\bullet$ If $p \geq q \geq 2$ then $n_C = 2$ and $n_B = 4$. 
The possible anomalies are $A_i$ with $1 \leq i \leq 6$ and 
$A_7$ iff $p \neq q$. Hence, $n_A = 7$ if $p \neq q$, $n_A = 6$ if
$p=q$, and either way GW fails. 
\hfill \break
$\bullet$ If $p \geq q =1$, the target manifolds are projective spaces $CP^p$
with $n_C = 1$ and $n_B = 3$.
If in addition $p \geq 2$, then $A_2$, $A_3$ and $A_7$ are 
independent, but $A_4$, $A_5$ and $A_6$ can be expressed in terms of
them, so $n_A = 4$ and GW fails.
If $p=q=1$, however, then $A_7$ also reduces 
to a combination of $A_2$ and $A_3$, so $n_A= n_B = 3$ and GW works; 
this is the case $SU(2)/U(1) = CP^1 = S^2$.

We need not consider $\varepsilon$ tensors for the complex
Grassmannians because the field $\Phi$ transforms under $SU(p{+}q)$
according to (\ref{sym}a), {i.e.}~in the tensor product of the 
defining representation and its conjugate.
We need $\varepsilon$ tensors associated to each of these 
representations (one with indices up and one with indices down in 
traditional notation) to construct an invariant, but the
product of two such tensors reduces to a combination of $\delta$
tensors, and hence to traces.

\noindent
\emph{Spin-4 for quaternionic/symplectic Grassmannians}
\hfil \break
$\bullet$ If $p \geq q \geq 2$ then $n_C = 2$ and $n_B = 4$.
The counting is exactly like the real case, in keeping with the 
remarks following (\ref{anti}).
Hence GW works for $p = q$ but fails for $p \neq q$.
\hfill \break
$\bullet$ If $p \geq q =1$ then $n_C = 1$ and $n_B = 3$.
The anomalies behave more subtlety here.
For $p \geq 2$, one combination of $A_4$, $A_5$ and $A_7$ 
remains independent of $A_2$ and $A_3$, so $n_A = 4$ and GW 
fails. But for $p=q=1$ we find $A_4$, $A_5$ and $A_7$ all 
reduce to $A_2$ and $A_3$, so $n_A=3$ and GW works; this is the case 
$Sp(2)/Sp(1){\times}Sp(1) = SO(5)/SO(4) = S^4$.

\vspace{20pt}
\centerline{\bf 5. Details for the remaining models}

The remaining families of classical symmetric spaces are
\be
SU(n)/SO(n) \ , \qquad 
SU(2n)/Sp(n) \ , \qquad 
SO(2n)/U(n) \ , \qquad 
Sp(n)/U(n) 
\nonumber \ee
The first two and last two sequences are very similar in character.

For $SU(n)/SO(n)$: $\Phi$ is an $n{\times}n$ complex matrix
and (\ref{sym}b,\ref{form}b,\ref{hinv}b) hold with
\be 
\Phi^T =\Phi \ , \qquad \Phi \Phi^* =1 \ , \qquad \det \Phi = 1 \ , \qquad
N =1 \ .
\ee
There is a symmetry
$ \mu : \Phi \mapsto M \Phi M^T$ where $M$ is a unitary
matrix with $\det M = -1$. Trace-type invariants (\ref{ginv}b), and
hence all conserved quantities, anomalies and derivatives, are
even under $\mu$. Any $SU(n)$-invariant constructed using an 
$\varepsilon$ tensor is odd under $\mu$, however, so we disregard
these.

For $SU(2n)/Sp(n)$: $\Phi$ is a $2n{\times}2n$ complex matrix 
and (\ref{sym}b,\ref{form}b,\ref{hinv}b) hold with
\be 
\Phi^T = - \Phi \ , \qquad \Phi \Phi^* = - 1 \ , \qquad {\rm Pf} \, \Phi
= 1 \ ,
\qquad N = J \ ,
\ee
where $J$ is a symplectic structure chosen to have ${\rm Pf} J = 1$.
The condition ${\rm Pf} \Phi = 1$ implies 
\be
\Phi_{[a_1 b_1} \Phi_{a_2 b_2} \ldots \Phi_{a_n b_n]} \ = \ 
\frac{2^n n!}{(2n)!} \,
\varepsilon_{a_1 b_1 a_2 b_2 \ldots a_n b_n}
\label{epsredux}\ee
and hence any $\varepsilon$ invariant can be re-expressed in terms
of traces. 

In these two families of sigma models the anomalies and derivatives can 
be written entirely in terms of $j$'s (because the $G$-invariants are of type 
(\ref{ginv}b)), so we can restrict attention to the lists 
(\ref{basicA}) and (\ref{basicB}). 
We noted in section 3 that this was sufficient to show 
quantum conservation of the spin-2 (energy-momentum) and spin-3 currents,
the latter existing when $n \geq 3$ (so that there is 3rd-order
invariant \cite{Evans00}). 
For the spin-4 currents, the counting is identical for 
members of either family with a common value of $n$. In fact, 
exactly the same anomalies and derivatives occur for the 
spin-4 current in the $SU(n)$ PCM, to which the same 
lists (\ref{basicA}) and (\ref{basicB}) apply \cite{PCM}.

\noindent 
\emph{Spin-4 in} $SU(n)/SO(n)$, $SU(2n)/Sp(n)$ \emph{and} $SU(n)$ (PCM)
\hfill \break
$\bullet$ If $n \geq 4$ then $n_C =2$, $n_B = 4$. 
The anomalies are $A_i$ in (\ref{basicA}): $n_A = 5$ and GW works.
\hfill \break
$\bullet$ If $n = 3$ then $n_C = 1$, $n_B = 3$.
One combination of $A_4$ and $A_5$ reduces to $A_2$ and $A_3$, but 
any distinct combination remains independent, so $n_A = 4$ and GW fails.
\hfill \break
$\bullet$ If $n= 2$ then $n_C =1$, $n_B=3$.
Now $A_4$ and $A_5$ each reduce to combinations of 
$A_2$ and $A_3$, so $n_A = 3$ and GW works. These target spaces 
are $SU(2)/SO(2) = S^2$, $SU(4)/Sp(2) = SO(6)/SO(5) = S^5$ and
$SU(2) = S^3$.
\hfill \break
(Further details relevant to the behaviour of these invariants 
are given in the appendix).
 
For $SO(2n)/U(n)$: $\Phi$ is a real, antisymmetric $2n{\times}2n$ matrix
and (\ref{sym},\ref{form},\ref{hinv}) hold with
\be 
\Phi^* = - \Phi^T = \Phi \ , \qquad \Phi^2 = - 1 \ , \qquad {\rm Pf} \, \Phi =
1 \ , \qquad N = J \ ,
\ee
where, again, $J$ is a symplectic structure with ${\rm Pf} J = 1$.
When $n$ is even there is a symmetry 
$\tau : \Phi \mapsto - \Phi $; when $n$ is odd there is a symmetry
$\tau : \Phi \mapsto - R \Phi R^T$ where 
$R$ is orthogonal, $\det R = -1$, and $RJ = -JR$ 
(this ensures ${\rm Pf} \, \Phi = 1$ is preserved); the currents 
(\ref{curr}) are unchanged by $\tau$.
Once again we may disregard $\varepsilon$ invariants because
(\ref{epsredux}) holds.

For $Sp(n)/U(n)$: $\Phi$ is an anti-hermitian $2n{\times}2n$ matrix
and (\ref{sym}a,\ref{form}a,\ref{hinv}a) hold with
\be 
\Phi^* = -\Phi^T = J \Phi J^{-1} \ , \qquad \Phi^2 = - 1 \ , \qquad 
\det \, \Phi = 1 \ , \qquad N = J \ .
\ee
There is a discrete symmetry $\tau : \Phi \mapsto -\Phi$ under which 
the higher-spin currents are inert.

For these last two families the field $\Phi$ is antihermitian
and this implies (\ref{anti}) again (compare with the 
Grassmannians, for which $\Phi$ is hermitian). Consequently:
(i) The trace of any odd number of currents vanishes, with or without
a $\Phi$ term. (ii) ${\rm Tr} (\Phi j^2) = 0$ for any $j$, from 
cyclicity of the trace. (iii) ${\rm Tr} (\Phi j j')$, which need not
vanish if $j \neq j'$, is even under (\ref{cc}) and odd under $\tau$.

\noindent
\emph{Spin-4 in} $SO(2n)/U(n)$ and $Sp(n)/U(n)$
\hfill \break
$\bullet$ If $n \geq 4$ in the first family, or $n \geq 2$ in the
second family, then $n_C =2$ and $n_B = 4$.
The anomalies are $A_i$ with $1 \leq i \leq 6$, $A_7$ being ruled out
by $\tau$, so $n_A = 6$ and GW fails.
\hfill \break
$\bullet$ The remaining cases are models we have already considered:
\hfill \break
$SO(6)/U(3) = SU(4)/S( \, U(3){\times}U(1) \,) = CP^3$ and
$SO(4)/U(2) = Sp(1)/U(1) = S^2$.

\vspace{15pt}
\centerline{\bf 6. The co-existence of local and non-local
quantum charges}

Returning to our results, stated at the end of section 3,
points (i) and (ii) imply that the integrable symmetric space
models (1) always possess local (as well as non-local) quantum conserved
charges. To discuss the significance of this, let us recall 
the situation for a PCM based on $G$.
In this model the non-local charges extend the Lie algebra $\g$ of $G$
to a quantum group structure, known as a Yangian $Y(\g) \supset \g$
(see e.g.~\cite{Bernard,MacKay92})
and the particle multiplets $V_i$ are irreducible representations
(irreps) of $Y(\g)$ \cite{ORW,PCM} 
(actually there are `left' and `right' copies of
this symmetry in the PCM, but this is not important here).
Let us denote by $(\mu)$ the $\g$-irrep with 
highest weight $\mu$, and let $\lambda_i$ be the 
fundamental weights dual to the simple roots and thus associated to the
nodes of the $\g$ Dynkin diagram. The PCM particle multiplets 
are $Y(\g)$-irreps in one-to-one correspondence with these nodes.
For the $A$ $(SU)$ and $C$ $(Sp)$
families, $V_i = (\lambda_i)$, while for the $B/D$ $(SO)$ families the 
$Y(\g)$-irreps are not $\g$-irreps in general, but they 
take the form $V_i = (\lambda_i) \oplus \{ \mbox{more} \}$.

One can, therefore, view the masses and $3$-point couplings or fusings 
in the PCM S-matrices \cite{ORW}
as arising from properties of Yangian (non-local charge)
representations and their tensor products. 
What is remarkable is that the same fusings and masses 
occur in affine Toda theories \cite{Corrigan},
where they follow from the existence of a set of quantum, 
conserved, commuting,
\emph{local} charges, with a particular set of spins equal to the 
exponents of $\g$ modulo its Coxeter number $h(\g)$
\cite{dorey91}. Actually, this statement requires careful
interpretation in the non-simply-laced cases, since these involve 
a \emph{restricted} fusing rule 
based on a dual pair of Kac-Moody algebras \cite{dorey93,chari95}.
Now, it was shown in \cite{PCM} that the PCMs possess 
\emph{classical}, commuting, conserved local charges with precisely 
this same set of spins (see \cite{Evans01} for the exceptional groups)
and, as described in section 3, some of them survive 
quantization. It is natural 
to conjecture \cite{PCM} that all these commuting charges 
survive, thereby fitting very nicely with the Yangian structure.

\[
\begin{array}{|c|c|c|}
\hline
G & \mathrm{exponents} & h(\g) \\
\hline 
&& \\
A_n = SU(n{+}1) & 1, 2, \ldots , n  & n{+}1 \\[0.1in]
B_n = SO(2n{+}1)  & 1, 3, \ldots, 2n{-}1 & 2n \\[0.1in]
C_n = Sp(n) & 1, 3, \ldots , 2n{-}1 & 2n \\[0.1in]
D_n = SO(2n) & 1, 3, \ldots , 2n{-}3; n{-}1 & 2n{-}2\\[0.05in]
\hline
\end{array}
\]

Exactly the same ideas can now be applied to the $G/H$ symmetric space 
sigma models. Here too the quantum non-local charges \cite{Luscher,AFG}
(see also \cite{EF,AABZ})
generate a Yangian, $Y(\g) \supset \g$, and the $S$-matrix
constructions of \cite{Fendley:2000,Babichenko} can be interpreted in
these terms, although the $Y(\g)$ irreps differ from
those in the PCMs (see below). 
It is also known \cite{Evans00} (and \cite{Evans01} for
the exceptional cases) that the $G/H$ model possesses
classical, commuting, conserved, local charges, with spins equal to
the exponents of $K$, a certain Lie group/algebra which should be 
regarded as encoding the simple root structure of $G/H$ \cite{Helgason}.
The obvious question is whether all these commuting charges could
survive quantization, and in particular whether this would be
consistent with the 
$S$-matrices proposed in \cite{Fendley:2000} and \cite{Babichenko}.

\[
\begin{array}{|c|c|c|}
\hline
G/H & K & V_i \\
\hline&  & \\
S^n = SO(n{+}1)/SO(n) & A_1 & (\lambda_1) \\[0.1in]
SU(2n{+}2)/Sp(n{+}1) & A_n = SU(n{+}1)
& (\lambda_{2i}) \\[0.1in]
Sp(2n)/Sp(n){\times}Sp(n) & C_n = Sp(n)&
(\lambda_{2i}) \\[0.1in]
SU(n{+}1)/SO(n{+}1) & A_{n} = SU(n{+}1) & 
(2\lambda_i) \\[0.1in]
SO(2n)/SO(n){\times}SO(n) & D_n = SO(2n) & (2\lambda_i) \oplus \{
{\mathrm{more}} \}\\[0.05in]
\hline
\end{array}
\]

For a basic test of this suggestion, consider the more detailed conclusions
(iii), (iv) and also (v), at the end of section 3.
Noting that a conserved charge of spin $s$ corresponds to a conserved 
current of spin $s{+}1$, we see that, for those models which are 
actually quantum integrable, a classical conservation law with $s=2$ 
always survives. We have shown that conservation laws with $s=3$ 
also survive quantization in these theories,
with the possible exception of the $SU(3)$ PCM and the $SU(3)/SO(3)$ and 
$SU(6)/Sp(3)$ sigma-models, for which GW fails.
But these are exactly the integrable examples 
for which \emph{there is no charge of spin 3 in the classically
commuting set constructed in} \cite{PCM, Evans00}, as can be checked
from the data in the tables. 
Thus the GW analysis is precisely compatible with the claim
that the classical \emph{commuting} charges generalise to the quantum theory.

Much more stringent consistency requirements 
follow by taking the fusings that would be predicted by the
quantum local charges, according to Dorey's rule
\cite{dorey91}, and comparing these 
with the behaviour of the 
Yangian representations $V_i$ given in the table.
We have included the $S^n$ models for completeness; these have just a
single irrep which is the vector of $\frak{so}(n{+}1)$ \cite{ZZ}.
For the remaining families, the representations $V_i$ with 
$i = 1 , \ldots , n$ are those proposed in the S-matrix constructions 
of \cite{Fendley:2000,Babichenko}.
Note that $(2 \lambda_i)$ appears in the symmetrized product of 
the $(\lambda_i)$ with itself.
The $SO(2n)/SO(n){\times}SO(n)$ case is by far the most
complicated, because in general the $Y(\frak{so}(2n))$-irreps are not
$\frak{so}(2n)$-irreps. However, those built from the vector ($i=1$) 
and spinor ($i=n{-}1, \, n$) representations are:
$V_i  = (2\lambda_i )$, for $i = 1 , n{-}1, n$.\footnote{ 
The first few higher tensor multiplets
after $V_1$ are
$
V_2\equiv(2\lambda_2)\oplus(\lambda_2)\oplus(0)$, 
$(2\lambda_3)\oplus(\lambda_3+\lambda_1)\oplus V_1$,
$(2\lambda_4)\oplus(\lambda_4+\lambda_2)\oplus(\lambda_4)\oplus
V_2$; the full formula is given in \cite{kleber}.}

For three of the families it is relatively straightforward to
compute the three-point-couplings. For $SU(2n{+}2)/Sp(n{+}1)$ and 
$Sp(2n)/Sp(n){\times}Sp(n)$ the multiplets are just a subset of the 
fundamental $Y(\g)$-irreps (the even-labelled ones), 
which are also $\g$-irreps, and so their tensor products and
associated S-matrix structure is fully understood. 
It is clear from \cite{Babichenko} and references
therein that the fusings are indeed those of the $A_n$ and restricted
$C_n$ cases respectively, as would be expected from Dorey's rule
\cite{dorey91,chari95}.
In the latter case, however, the masses are those
of the $A_{2n}^{(2)}$ Toda theory \cite{twist} (for which the
$R$-matrix fusings are the same as $C_n$); it would be interesting to 
investigate this case in more detail (similar subtleties arise in 
comparing the non-simply-laced Toda models and PCMs \cite{PCM}).
For $SU(n{+}1)/SO(n{+}1)$ the irreps $V_i$ are still $\g$-irreps,
but they are not fundamental, so some work must be done.
Using the tensor product graph (TPG) \cite{TPG},
one can show that $V_j = (2\lambda_j)$ and $V_k= (2\lambda_k)$ fuse
to give, for example, $V_{j+k} = (2\lambda_{j+k})$ 
at $\theta = i \pi (j{+}k) / (n{+}1)$ if $j+k < n{+}1$. In general 
we find exactly the same fusings and rapidity differences as for 
the irreps $(\lambda_j)$ and $(\lambda_k)$ 
with the original, undoubled, weights. The couplings and masses are 
therefore those of $A_n$, in accordance, once again, with 
expectations based on quantum local charges.

The $SO(2n)/SO(n)\times SO(n)$ case is, of course, harder, and 
we are unable to carry out all the necessary calculations because 
the TPG cannot handle $\g$-reducible representations. However, one can compute 
the TPGs for the irreps $V_1$, $V_{n-1}$ and $V_n$ 
with one another ($V_1 \otimes V_1$ is given in \cite{Fendley:2000,TPG})
and it is remarkable that once again the same couplings occur, and at
the same rapidity differences, as for 
$(\lambda_1)$, $(\lambda_{n-1})$ and $(\lambda_n)$ in $D_n$. It is natural to 
expect, therefore, that the same applies for all the $V_i$ and that the
masses and couplings are just those of $D_n$. 

Given our findings, it would be interesting to investigate 
whether the central result of 
\cite{chari95}---on the Dorey's-rule structure of tensor products 
of fundamental $Y(\g)$ representations---holds more generally when the 
highest weights
of the representations are doubled, or perhaps even multiplied 
by some integer. Another very natural question is whether there is any 
deeper significance to the root system $K$ from which one could 
gain a more direct understanding of the multiplet structure in the 
quantum $G/H$ model.
\vspace{20pt}

{\bf Acknowledgments}:
The research of JME is supported in part by Gonville and Caius
College, Cambridge. CASY is supported by a PPARC studentship. DK is
grateful for a NSF graduate research fellowship and a Columbia
University Euretta J.~Kellett fellowship. 
This work was also supported by the EU `EUCLID' network,
ref.~HPRN-CT-2002-00325.
\vspace{30pt}

\centerline{\bf Appendix: more on certain 4th-order invariants}

To apply the GW counting argument, we need to know which of the 
anomaly terms $A_i$ with $2 \leq i \leq 7$ in (\ref{basicA}) and 
(\ref{extraA}) are independent when there is no primitive 4th-order 
invariant. 

\noindent
$\bullet$ 
\emph{$SU(n)$, $SU(n)/SO(n)$ and $SU(2n)/Sp(n)$}:
These spaces have a primitive symmetric 4th-order invariant when $n \geq
4$ \cite{Evans00} and the anomalies 
$A_i$ with $2 \leq i \leq 5$ in (\ref{basicA}) are then independent. 
When $n < 4$ there is no primitive 4th-order invariant and so one
combination of $A_4$ and $A_5$ can be expressed in terms of 
$A_2$ and $A_3$. To see whether or not there is still one combination 
of $A_4$ and $A_5$ which is independent of $A_2$ and $A_3$ it suffices,
by linearity, to determine whether or not
\be 
{\rm Tr} (X^2 Y^2) \ , \qquad {\rm Tr} (X^2) {\rm Tr} (Y^2) \ , \qquad
({\rm Tr} XY)^2
\label{indep}\ee
are independent for $X, Y \in \m$ (where, for $SU(n)$, $\m =
\frak{su}(n)$, its Lie algebra). 
\hfill \break
\phantom{X}
It is easy to check that 
${\rm Tr} X^4 = {\textstyle {\frac{1}{2}}} ({\rm Tr} X^2)^2$
for $X \in \m = \frak{su}(2)$ or $\frak{su}(3)$ and that 
${\rm Tr} ( X^2 Y^2 ) = {\textstyle {\frac{1}{2}}} ({\rm Tr} X^2)
({\rm Tr} Y^2)$ for $X, Y \in \frak{su}(2)$ ( or, in a different
representation, ${\rm Tr} ( X^2 Y^2 ) = {\textstyle {\frac{1}{4}}} 
({\rm Tr} X^2) ({\rm Tr} Y^2)
+ {\textstyle {\frac{1}{4}}} ({\rm Tr} XY)^2$ for 
$X, Y \in \frak{so}(3)$ ).
But the quantities (\ref{indep}) are 
independent for $X , Y \in \frak{su}(3)$, as may be checked by
judicious choices of these matrices.
\hfill \break
\phantom{X}
The symmetric spaces $SU(n)/SO(n)$ and $SU(2n)/Sp(n)$ behave 
very similarly. 
For the first family, $\m \subset \frak{su}(n)$,
consisting of symmetric, imaginary matrices.
For the second family, $\m$ consists of matrices of the block form 
{\small $\bmx x & y \\ y^* & -x^* \emx$}~with 
$x \in \frak{su}(n)$ and $y$ complex 
and $n{\times}n$ with $y^T = -y$. 
In both cases the quantities (\ref{indep}) are easily seen to be 
independent for $n=3$, but for $n=2$ standard properties 
of $su(2)$ matrices imply that $X^2$ and $Y^2$ are multiples of 
the identity matrix and hence the first two expressions in 
(\ref{indep}) are proportional.

\noindent
$\bullet$ \emph{Grassmannians}: 
For $X, Y \in \m$ we have the block forms 
\be
X = \bmx 0 & x \\  - x^\dagger & 0  \emx 
\ , \qquad
Y = \bmx 0 & y \\  - y^\dagger & 0  \emx 
\ , \qquad
N = \bmx 1 & 0  \\  0 & - 1  \emx 
\ , \qquad
\ee
where $x, y$ are $p{\times}q$ real or complex matrices, 
or in the quaternionic case $2p{\times}2q$ complex
matrices obeying a symplectic reality condition.
Using the fact that $g^{-1} \Phi g = N$ with $g^{-1} j g \in \m$ for each
current $j$, any relations among 
$A_i$ with $2 \leq i \leq 7$ in (\ref{basicA}) and (\ref{extraA}) are 
equivalent to relations among:
\bea
{\rm Tr} X^2 \, {\rm Tr} Y^2 & = & 4 \, {\rm Tr} \, x^\dagger x \, 
{\rm Tr} \, y^\dagger y
\nonumber\\
{\rm Tr} X Y \, {\rm Tr} X Y & = & 
( \, {\rm Tr} \, x^\dagger y \, + \, {\rm Tr} \, y^\dagger x \, )^2
\nonumber\\
{\rm Tr} \, NXY \, {\rm Tr} \, NXY & = & 
( \, {\rm Tr} \, x^\dagger y \, - \, {\rm Tr} \, y^\dagger x \, )^2 
\nonumber\\
{\rm Tr} \, XYXY & = & 
{\rm Tr} \, x y^\dagger x y^\dagger  
+ {\rm Tr} \, x^\dagger y x^\dagger y 
\nonumber\\
{\rm Tr} \, X^2 Y^2 & = & 
{\rm Tr} \, x x^\dagger y y^\dagger  
+ {\rm Tr} \, x^\dagger x y^\dagger y  
\nonumber
\\
{\rm Tr} (N X^2 Y^2) & = & 
{\rm Tr} \, x x^\dagger y y^\dagger  
- {\rm Tr} \, x^\dagger x y^\dagger y  
\eea
These formulas can be used to check various statements made in
section 5. In particular, in the real and complex cases with $q=1$, 
$x$ and $y$ are $p$-component column vectors. In the real case, 
with $p \geq 2$, there are clearly 2 invariants $x^2 y^2$ and $(x^T y)^2$.
In the complex case there are 3 independent combinations if $p \geq
2$, namely $(x^\dagger x) (y^\dagger y)$,
$(x^\dagger y) (y^\dagger x)$ and
$( \, x^\dagger y + y^\dagger x \,)^2$; but 
these reduce to 2 if $p=1$, namely 
$(x^* x) (y^* y)$ and $(x^* y + y^* x)^2$.
In the quaternionic case $x$ and $y$ are $2p{\times}2$ matrices, even 
when $q=1$. The independent combinations are 
$(\, {\rm Tr} \, x^\dagger x \, )(\, {\rm Tr} \, y^\dagger y \,)$,
$(\, {\rm Tr} \, x^\dagger y \, )(\, {\rm Tr} \, y^\dagger x \,)$
and a third, ${\rm Tr} (x^\dagger y x^\dagger y ) =
{\rm Tr} (y^\dagger x y^\dagger x)$ which is independent if $p \geq
2$, but not if $p=1$.

 \end{document}